\documentclass[aps,pra,superscriptaddress,preprint,amsmath,amssymb,showpacs]{revtex4-1}
\usepackage[english]{babel}
\usepackage{amssymb,bm}
\usepackage{graphicx}
\usepackage{amsmath}
\usepackage[colorlinks=true,citecolor=blue,urlcolor=blue]{hyperref}
\begin{document}

\title{Spin dynamics of a hydrogen atom in a periodic magnetic structure}
\author{A.I. Milstein}
\email{A.I.Milstein@inp.nsk.su}
\affiliation{Budker Institute of Nuclear Physics of SB RAS, 630090 Novosibirsk, Russia}
\author{Yu.V. Shestakov}
\email{Yu.V.Shestakov@inp.nsk.su}
\affiliation{Budker Institute of Nuclear Physics of SB RAS, 630090 Novosibirsk, Russia}
\author{D.K. Toporkov}
\email{D.K.Toporkov@inp.nsk.su}
\affiliation{Budker Institute of Nuclear Physics of SB RAS, 630090 Novosibirsk, Russia}

\date{\today}

\begin{abstract}
The spin dynamics of a hydrogen atom during the passage of a periodic magnetic structure is discussed. The occupation numbers of the components of the hyperfine structure are considered as a function of time. The characteristic low-frequency oscillations are visible, which have a direct analogue in the effect of nuclear magnetic resonance. An envelope forms  of these  oscillations are found using the Krylov-Bogolyubov-Mitropol'skii  method. The dependence of spin dynamics on the parameters of the magnetic structure is investigated. It is shown  that this dependence is very sensitive to the structure of the magnetic field. 
\end{abstract}

\maketitle

\section{Introduction} 
At present, there are quite a few works in which the spin dynamics of atoms and molecules are studied during the passage of various magnetic structures, see, e.g. \cite {R1956, C2009, S1999, E2015, S1967, T2013} and the references  therein. The study of this dynamics is important, first of all, from a practical point of view. There are various interpretations of the obtained experimental results. Using the spin dynamics of a hydrogen atom passing through a periodic magnetic structure as an example, we show that in this problem there is a direct analogy between the time dependence of the level population and the phenomenon of nuclear magnetic resonance. We restrict ourselves to the simplest case of an electron in the $ 1s $ state, taking into account the hyperfine interaction of the electron and the nucleus.

Consider the magnetic moment $ \bm {\mu} $ of a particle with spin $ S = 1 $ in a magnetic field, which is a superposition of a constant magnetic field $ H_0 $ directed along the $ z $ axis and a field $ H_1 $ rotating with a frequency $ \omega $ in the $ xy $ plane. Choosing the $ z $ axis as the spin quantization axis, it is easy to calculate from the Pauli equation  \cite{LL} the probabilities $ w_1 $, $ w_0 $, and $ w_{- 1} $  to find a particle in the state with the corresponding projection $ S_z $, if $ w_1 = 1 $ and $ w_0=w_{-1} = 0 $ in the initial  time. One has
 \begin{align}\label{w}
 &w_1=(1-\varkappa)^2\,,\quad w_0=2\varkappa(1-\varkappa)\,,\quad w_{-1}=\varkappa^2\,,\nonumber\\
 &\varkappa=\dfrac{(\mu H_1)^2}{(\hbar\Omega)^2}\sin^2(\Omega t/2)\,,\quad \Omega=\dfrac{1}{\hbar}\sqrt{(\mu H_0-\hbar\omega)^2+(\mu H_1)^2}\,.
 \end{align}
 In resonance, $ \hbar\omega = \mu H_0 $. If $ H_1\ll H_0 $, then near the resonance $ \Omega\ll \omega $. If $ H_1 \gg H_0 $, then near the resonance $ \Omega \gg \omega $.
 
 Let us now consider a hydrogen atom in the $ 1s $ state, moving with velocity $ v $ in a time-independent  but non-uniform periodic magnetic field. Taking into account the hyperfine interaction of an electron and a proton, we have four states: three states with a total spin of electron and proton to be $S = 1 $  and projections $ S_z = 
 \pm1, \, 0 $, and one state with a total spin $ S= 0 $. Let us consider the magnetic field  much smaller than the magnetic field of the proton magnetic moment  at a distance $ a_B $ (Bohr radius), i.e. $ H \ll 100 \, \mbox{G} $. In this case, one can neglect the transitions between the states with $ S = 1 $ and $ S = 0 $. The simplest azimuthally symmetric configuration of a magnetic field periodic along $ z $ axis, satisfying the equation $ \mbox {div} H = 0 $ has the form
  \begin{equation}
 H_z=H_0\sin(kz)\,,\quad H_\rho=-\dfrac{1}{2}k\rho\,H_0\cos(kz)\,,
 \end{equation}
 where $H_0$ and $k$ are some constants. Passing to the rest frame of the hydrogen atom, we have a time-dependent magnetic field
  \begin{equation}
 H_z(t)=H_0\sin(kvt)\,,\quad H_\rho=-\dfrac{1}{2}k\rho\,H_0\cos(kvt)\,.
 \end{equation}

The atomic magnetic moment operator is $ \bm {\mu}_H = 2 \mu_e \bm{s}_e +2 \mu_p \bm {s}_p $, where $\bm{s}_e$ and $\bm {s}_p$ are the electron and proton spin operators, $\mu_e$ and $\mu_p$ are their magnetic moments. The matrix element  of $\bm {\mu}_H$ over the states with a fixed $ \bm S = \bm s_e + \bm s_p $ coincides with the matrix element of the operator  $ \bm {\mu}_H = (\mu_e + \mu_p) \bm S \approx - \mu_B \bm S $, where $\mu_B$ is the Bohr magneton. Using  the variable $ \tau = kvt $, we write the Pauli equation \cite {LL}, which describes the spin dynamics, in the form 
\begin{align}\label{EP}
&i\frac{\partial}{\partial\,\tau}\psi=B[S_z\sin\tau-\sqrt{2}\lambda\,S_x\cos\tau]\psi\,,\quad B=\dfrac{\mu_BH_0}{\hbar kv}\,,\quad \lambda=\dfrac{k\rho}{2\sqrt{2}}\,,\quad \psi=
\left( \begin{array}{c} a_1 \\ a_2 \\ a_3 \end{array}\right)\,,
\end{align}
where $S_z$ and $S_x$ are matrices corresponding to the $ z $ and $ x $ components of the spin operator for $ S = 1 $. We rewrite the equation \eqref {EP} as follows
\begin{align}\label{EP1}
&i\,\dot{a_1}=-B[-\sin\tau\cdot a_1+\lambda\cos\tau\cdot a_2]\,,\nonumber\\
&i\,\dot{a_2}=-B\lambda\cos\tau\cdot(a_1+a_3)\,,\nonumber\\
&i\,\dot{a_3}=-B[\sin\tau\cdot a_3+\lambda\cos\tau\cdot a_2]\,,
\end{align}
where $\dot{a}_i=\partial a_i/\partial\tau$. 
For $ v = 1500 \, \mbox {m/s} $, $ H_0 = 1 \,\mbox {G} $, $ k = 2\pi \,\mbox {cm}^ {-1} $ we have $ B =9 $. Therefore, for the fields $ H_0 = 0.01 \div 0.1 \, \mbox {G} $  we obtain
$ B <1 $. In our work we  assume  $ \lambda <1 $. Our task is to analyze the solutions of the equations \eqref{EP1} for various  values of the parameters $ B $ and $ \lambda <1 $.

\section{Spin dynamics at  ${B \lesssim 1 }$ and ${B \gtrsim 1 }$ at $\lambda<1$.}

\begin{figure}[h]
	\centering
	\includegraphics[width=0.48\linewidth]{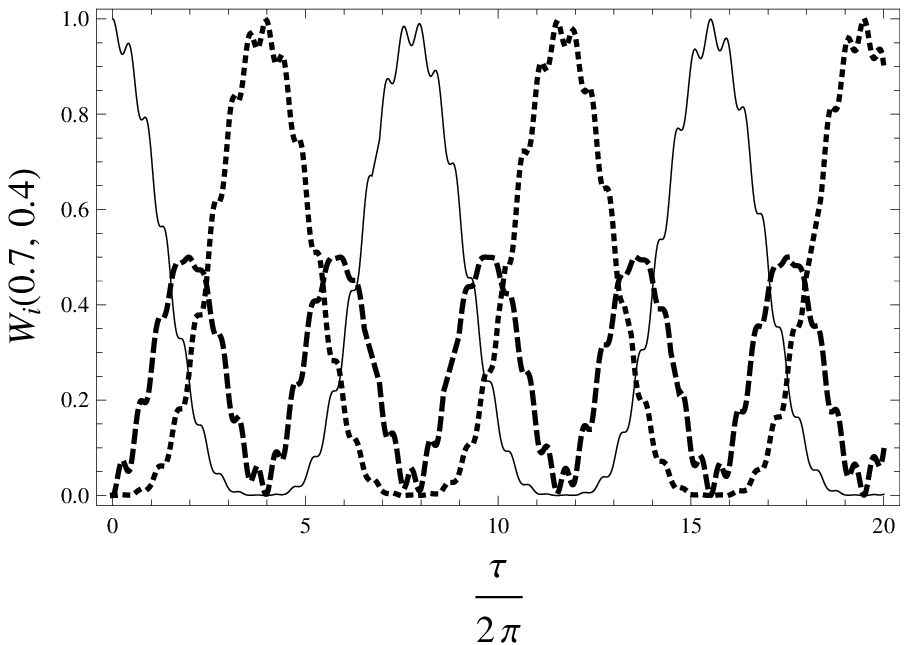}
	\includegraphics[width=0.48\linewidth]{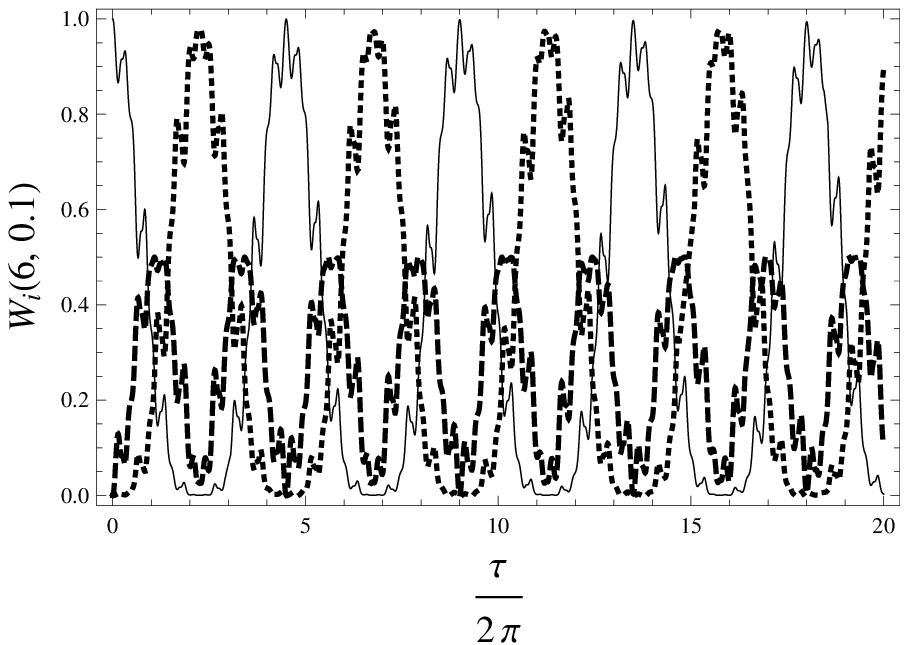}
		\caption{Dependence of $ W_i (B, \lambda) $ on $ \tau $ for various values of the parameters $ B $ and $ \lambda $. The solid line corresponds to the function $ W_1 $, the dashed line corresponds to the function $ W_0 $, and the dotted line corresponds to the function $ W_{- 1} $. }
	\label{fig1}
\end{figure}

Denote by $ W_1 (B, \lambda) = | a_1 | ^ 2 $, $ W_2 (B, \lambda) = | a_2 | ^ 2 $, and $ W_3 (B, \lambda) = | a_3 | ^ 2 $ the probabilities to have, respectively, spin projections $ 1 $, $ 0 $ and $ -1 $ at time $ \tau $.  A typical  dependence of $ W_i $ on $ \tau $ for  ${B \lesssim 1 }$ and ${B \gtrsim 1 }$ at  $\lambda<1$ is shown in Fig.~\ref{fig1}. We chose the boundary conditions $ a_1 = 1 $, $ a_2 = 0 $, and $ a_3 = 0 $ at $ \tau = 0 $. However, the characteristic behavior of the probabilities $ W_i $ does not depend much on the initial value of $ \tau $. Note that for $ \lambda = 0 $ the probabilities $ W_i $ are independent of $ \tau $.
It is seen from Fig.~\ref{fig1} that  the probabilities $ W_i $ are periodic functions of $ \tau $, and for the oscillation period $ T $ we have $ T /(2 \pi ) \gg 1 $. Shallow ripples with a period  of a magnetic structure are superimposed on the smooth envelopes of probabilities. The observed picture fully corresponds to the time dependence of polarization described above in the case of nuclear magnetic resonance. We checked that a specific form of the periodic dependence of $ H_z $ and  $ H_ \rho $ does not change the qualitative picture of spin dynamics.

\begin{figure}[h]
	\centering
	\includegraphics[width=0.48\linewidth]{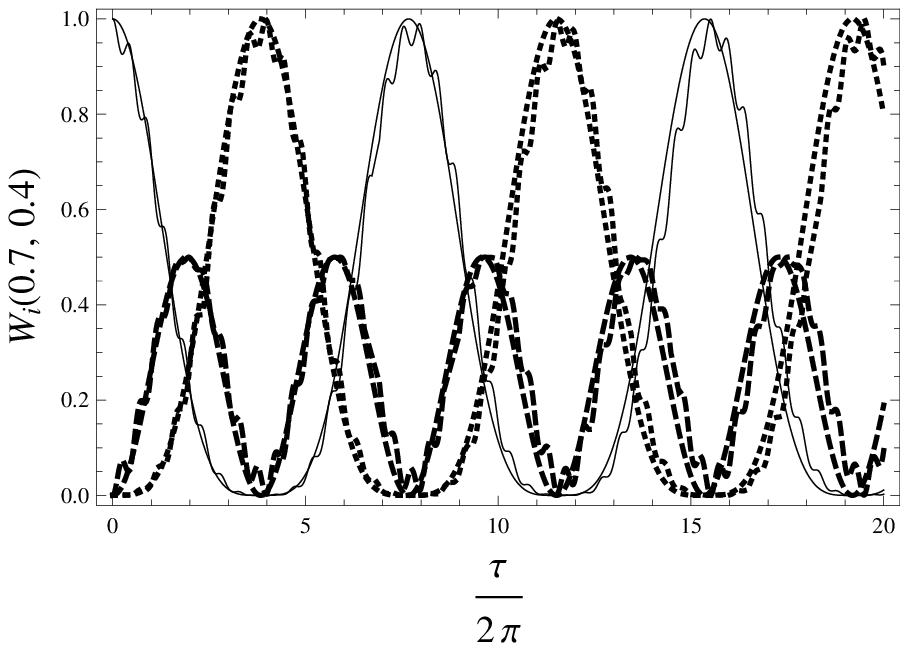}
	\includegraphics[width=0.48\linewidth]{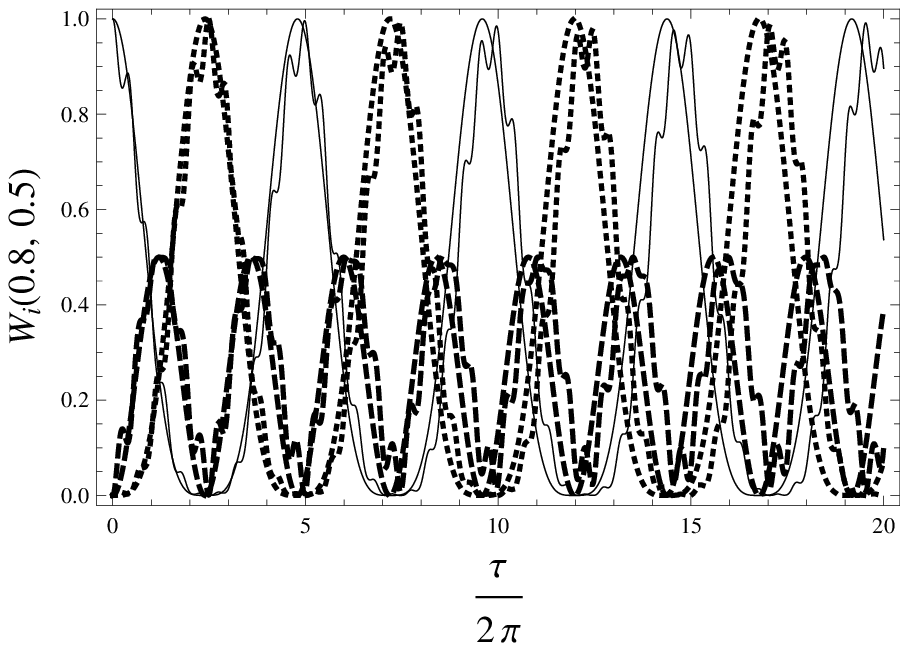}
		\caption{ Comparison of the functions $ W_i(B, \lambda) $ and $ \overline {W}_i(\Omega_0) $ for a few values of $ B <1 $ and $ \lambda <1 $. The solid lines correspond to the functions $ W_1 $ and $ \overline {W}_1 $, the dashed lines correspond to the functions $ W_0 $ and $ \overline {W}_0 $ and the dotted lines correspond to the functions $ W_{- 1} $ and $ \overline {W}_{- 1} $. Lines with ripples correspond to the functions $ W_i $, and smooth lines correspond to the functions $ \overline {W}_i $.}
	\label{fig2}
\end{figure}

Since $ T/(2\pi) \gg 1 $, it is possible to find the envelope form $ \overline {W} _i $ of the  probabilities $ W_i $ using the Krylov-Bogolyubov-Mitropol'skii  method \cite {BM} of averaging over high-frequency oscillations. To do this, we write the amplitudes $a_i$ in Eq.~\eqref {EP1} as $a_i=|a_i|\exp({-i\phi_i})$,  $|a_i|=X_i+x_i$, and $\phi_i=\Phi_i+\varphi_i$, where $ X_i $ and $ \Phi_i $ correspond to low-frequency oscillations, and $ x_i $ and $ \varphi_i $ correspond to high-frequency oscillations. Then, in the leading approximation in   $ \lambda <1$ we obtain:
\begin{align}\label{EP2}
&\dot{X_1}=-\lambda B\,\overline{\cos\tau\sin\varphi_{12}}\, X_2\,,\nonumber\\
&\dot{X_2}=\lambda B[\,\overline{\cos\tau\sin\varphi_{12}}\,X_1+\overline{\cos\tau\sin\varphi_{32}}\,X_3]\,,
\nonumber\\
&\dot{X_3}=-\lambda B\,\overline{\cos\tau\sin\varphi_{32}}\,X_2\,,\nonumber\\
&\dot{\varphi}_{12}= B\sin\tau\,,\quad \dot{\varphi}_{32}=- B\sin\tau\,,\nonumber\\
&\varphi_{12}= \varphi_{1}- \varphi_{2}\,,\quad \varphi_{32}= \varphi_{3}- \varphi_{2}\,,
\end{align}
where $\overline{A}$ means the averaging over high-frequency oscillations. Thus, 
$$\varphi_{12}= -B\cos\tau\,,\quad \varphi_{32}= B\cos\tau\,.$$
Using the integral
$$ \dfrac{1}{2\pi}\int_{-\pi}^{\pi}\cos\tau\,\sin(B\cos\tau)\,d\tau=J_1(B)\,,$$
where $J_n(x)$ is the Bessel function,  we perform averaging over high-frequency oscillations in Eq.~\eqref{EP2} and obtain
\begin{align}\label{EP4}
&\dot{X_1}=G_0 X_2\,,\nonumber\\
&\dot{X_2}=G_0 [-X_1+X_3]\,,
\nonumber\\
&\dot{X_3}=-G_0\,X_2\,,
\end{align}
where $G_0=\lambda B\,J_1(B)$. Finally, we arrive at the probabilities $\overline{W}_i=X_i^2$:
\begin{align}\label{barW}
&\overline{W}_1=\cos^4(\Omega_0\tau/2)\,,\quad \overline{W}_0=2\sin^2(\Omega_0\tau/2)\cos^2(\Omega_0\tau/2)\,,\quad
\overline{W}_{-1}=\sin^4(\Omega_0\tau/2)\,,
\end{align}
where the oscillation frequency  $\Omega_0 $ reads
\begin{align}\label{omega01}
& \Omega_0=\sqrt{2}\lambda B\,J_1(B)\,.
\end{align}
This formula is valid at $\Omega_0\ll 1$. For $B\ll 1$ we have
$\Omega_0=\lambda B^2/\sqrt{2}\,$.
A comparison of $ W_i $ and $ \overline {W} _i $ is made in Fig.~\ref{fig2} for a few values of $ B <1 $ and $ \lambda <1 $.
One can see from Fig.~\ref {fig2} an excellent agreement between $ W_i $ and $ \overline {W}_i $. 

\begin{figure}[h]
	\centering
	\includegraphics[width=0.47\linewidth]{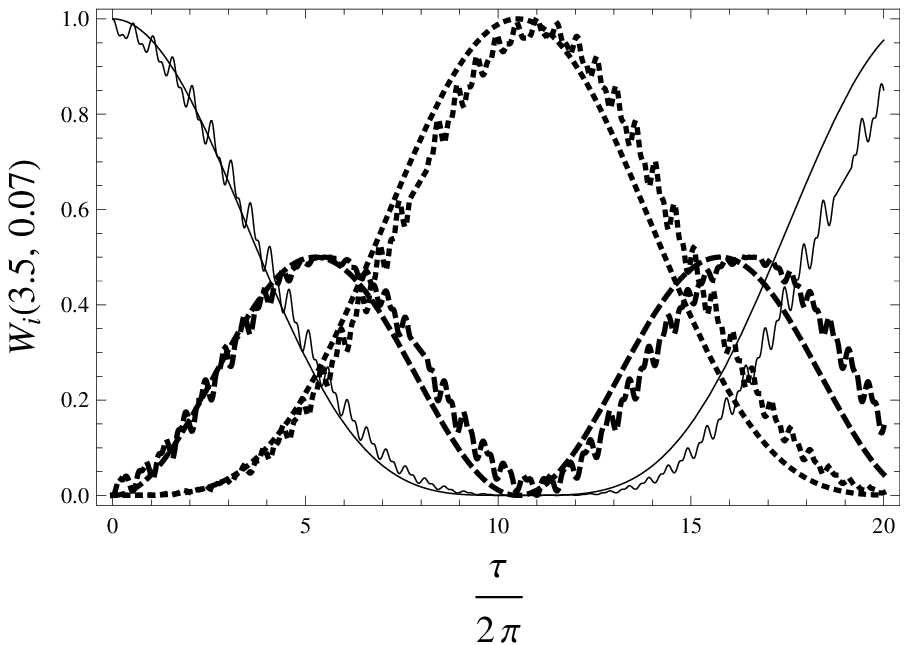}
	\includegraphics[width=0.47\linewidth]{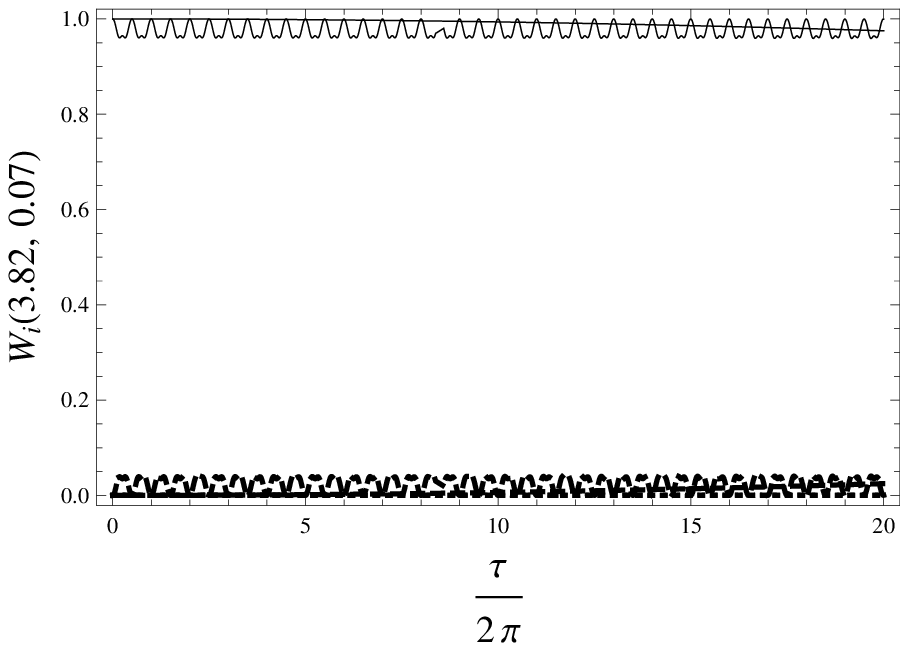}
	\includegraphics[width=0.47\linewidth]{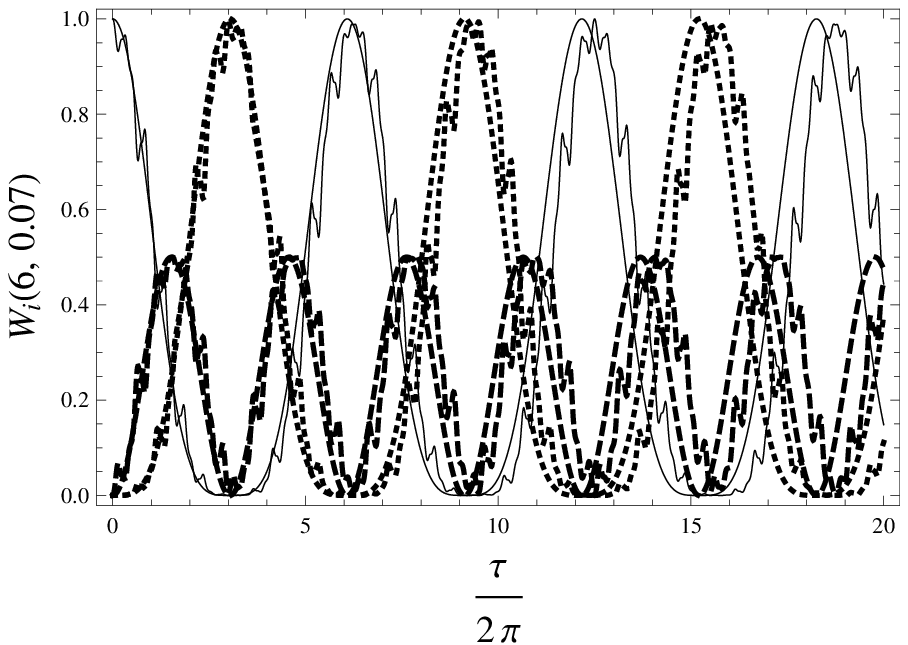}
	\includegraphics[width=0.47\linewidth]{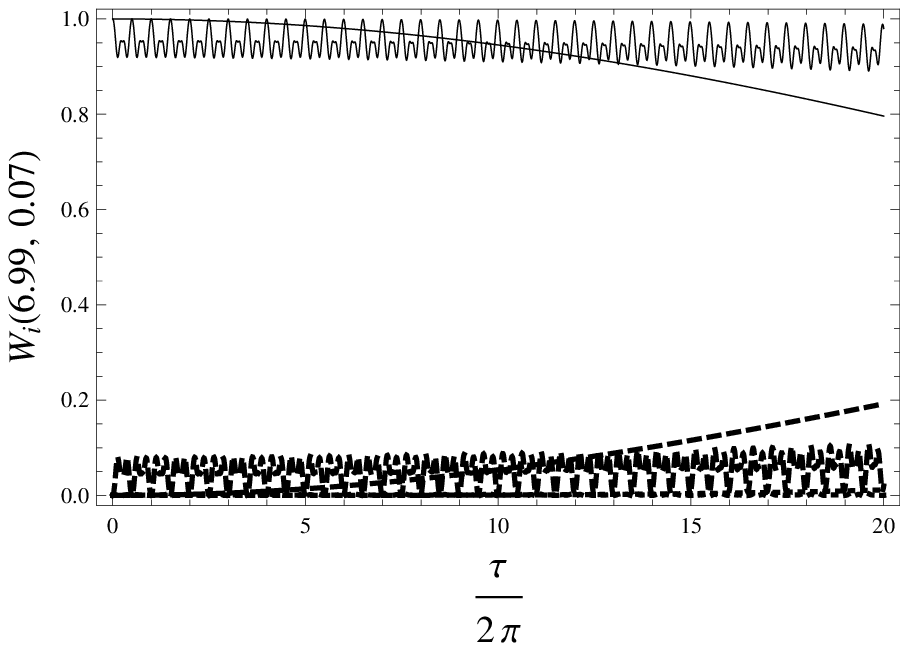}
	\caption{ Comparison of the functions $ W_i(B, \lambda) $ and $ \overline {W}_i(\Omega_0) $ for a few values of $ B >1 $ and $ \lambda=0.07$. The solid lines correspond to the functions $ W_1 $ and $ \overline {W}_1 $, the dashed lines correspond to the functions $ W_0 $ and $ \overline {W}_0 $, and the dotted lines correspond to the functions $ W_{- 1} $ and $ \overline {W}_{- 1} $. Lines with ripples correspond to the functions $ W_i $, and smooth lines correspond to the functions $ \overline {W}_i $.}
	\label{fig3}
\end{figure}

Let us consider the case ${B \gtrsim 1 }$ at $ \lambda <1 $.
Fig.~\ref{fig3} shows the dependence of $ W_i $ and $ \overline {W} _i $ on $\tau$ for a few values of $ B >1$ at $ \lambda =0.07 $. It is seen that this dependence is very sensitive to the value of $B$. In the vicinity of zeros of the Bessel function $J_1(B)$ the frequency  $\Omega_0 $ vanishes (these zeros are $B^*= 3.83,\, 7.02,\, 10.17,\, 13.32...$).

The expressions \eqref{barW} for $\overline{W}_i$ coincide with  \eqref {w} for $ w_i $ in the case of nuclear magnetic resonance at $\hbar\omega = \mu H_0 $ after the substitutions $\Omega_0\rightarrow \mu H_1$  and $\tau\rightarrow t$.

\begin{figure}[h]
	\centering
	\includegraphics[width=0.5\linewidth]{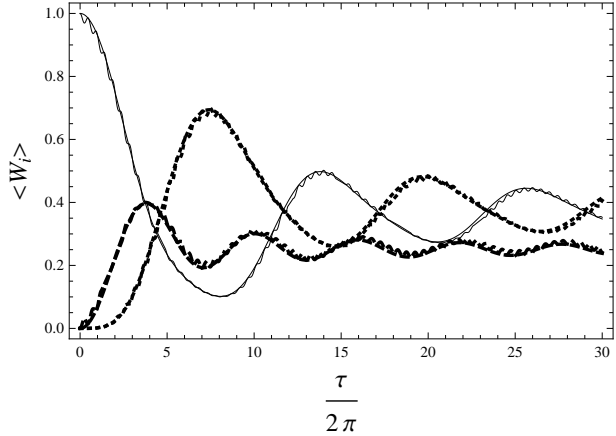}
	\caption{Dependence of the functions $\langle W_i \rangle$, Eq.~\eqref{av}, and $\langle\overline{W}_i\rangle$ on $ \tau $ for $ B = 0.5 $ and $ \lambda_0 = 0.5 $; the solid lines correspond to the functions $\langle W_1\rangle$  and $\langle\overline{W}_1\rangle$, the dashed lines correspond to the functions $\langle W_0\rangle$ and $\langle\overline{W}_0\rangle$, and the dotted lines correspond to the functions $\langle W_{-1}\rangle$ and $\langle\overline{W}_{-1}\rangle$. Lines with ripples correspond to the functions $\langle W_i \rangle$ and smooth lines correspond to the functions
		$\langle\overline{W}_i\rangle$.  }	
	\label{fig4}
\end{figure}

Above, we investigated the dependence of $ W_i $ on $ \tau $ for various values of $ B$ and $ \lambda <1 $. A fixed value of $ \lambda $ means a fixed value of the impact parameter $ \rho $. Since the frequency $ \Omega_0 $ of the envelopes $ W_i $ depends on $ \lambda $, averaging over impact parameters can distort the oscillation pattern. To elucidate this statement, let us  consider a beam with a transverse size $ \rho_0 $ and uniform density. Then the average value of $ \langle W_i \rangle $ of probabilities $ W_i $ is
\begin{equation}\label{av} 
\langle W_i\rangle=\dfrac{2}{\rho_0^2}\int_0^{\rho_0}W_i(B,\lambda)\rho\,d\rho\,=\dfrac{2}{\lambda_0^2}
\int_0^{\lambda_0}W_i(B,\lambda)\lambda\,d\lambda\,,\quad \lambda_0=\dfrac{k\rho_0}{2\sqrt{2}}\,.
\end{equation}
The dependence of $ \langle W_i\rangle $ on $ \tau $ for $ B = 0.5 $ and $ \lambda_0 = 0.5 $ is shown in Fig.~\ref{fig4}. For comparison, the same figure shows the envelopes obtained from Eqs.~\eqref{barW} and \eqref{av}.

As it follows  from Eqs.~\eqref{barW} and \eqref{av}, at $\tau\rightarrow\infty$  the oscillations in $\langle W_i \rangle$ fade out and $\langle W_1 \rangle\rightarrow 3/8$, $\langle W_2 \rangle\rightarrow 1/4$, and $\langle W_3 \rangle\rightarrow 3/8$ regardless of the values of $ B $ and $ \lambda_0 $.

\section{Conclusion}
It is shown that in a hydrogen atom moving along the $ z $ axis in a periodic magnetic field, the probabilities $ W_i $ to have certain projections of the total spin (electron and proton) on the $ z $ axis are also periodic functions of $ z $. The period of these functions is much larger than the period of the magnetic field. Using the Krylov-Bogolyubov-Mitropol'skii method, we found the envelopes of the functions $ W_i $ and showed that the oscillation period is a function of the amplitude and period of the magnetic field, the velocity of the atom, and the impact parameter of the atom relative to the axis of the magnetic system. We also showed that averaging over the impact parameter of atoms in the beam leads to oscillation damping. Therefore, to observe the effect of oscillations and to control the polarization using this effect, it is necessary to prepare a beam in which all atoms fly at the same distance from the axis of the magnetic system. We have demonstrated that the spin dynamics in a periodic magnetic system have a direct analogy with the effect of nuclear magnetic resonance. High sensitivity of the period of low-frequency
oscillations to values of the parameter $B$ may, in principle,  be used in applications.

{\bf Acknowledgements.}\\
We are grateful to R. Engels for useful  discussions.

\end{document}